# Analysis of Nonlinear Soil-Structure Interaction Effects on the response of Three-Dimensional Frame Structures using a One-Direction Three-Component Wave Propagation Model

M.P. Santisi d'Avila[1] and F. Lopez Caballero[2]

[1]Laboratoire J.A. Dieudonné
Université de Nice-Sophia Antipolis, France
[2]Laboratoire MSS-Mat, CentraleSupélec, Paris, France

**Abstract**

In this paper, a model of one-directional propagation of three-component seismic waves in a nonlinear multilayered soil profile is coupled with a multi-story multispan frame model to consider, in a simple way, the soil-structure interaction modelled in a finite element scheme. Modeling the three-component wave propagation enables the effects of a soil multiaxial stress state to be taken into account. These reduce soil strength and increase nonlinear effects, compared with the axial stress state. The simultaneous propagation of three components allows the prediction of the incident direction of seismic loading at the ground surface and the analysis of the behavior of a frame structure shaken by a three-component earthquake.
A parametric study is carried out to characterize the changes in the ground motion due to dynamic features of the structure, for different incident wavefield properties and soil nonlinear effects. A seismic response depending on parameters such as the frequency content of soil and structure and the polarization of seismic waves is observed.

## 1 Introduction

The seismic response of structures depends on ground motion features and on mechanical properties of structure and soil. The strong ground motion at the surface of a soil basin, shaking the base of a structure, is also influenced by the dynamic properties of the structure it-self. The stratigraphy of the soil profile and the mechanical features of the soil modify the seismic waves, propagating from the bedrock to the surface, and consequently the seismic loading at the base of structures. Furthermore, structure oscillation at the soil surface modifies the ground motion. This effect, known as soil-structure interaction (SSI), is influenced by the difference in principal frequencies of the soil and structure, due by differences in terms of mass and stiffness. The wave propagation of an earthquake along a soil profile, where a free surface is assumed at the top, does not allow take into account the effects of SSI (Saez et al. [13]).

The research, described in this paper, is directed towards the analysis of the behavior of a system composed of a frame structure over surface soil layers, under seismic loading. The one-dimensional (1D) multilayered soil profile is discretized by three-node line finite elements and a three-dimensional (3D) constitutive relationship describes the nonlinear soil behavior under cyclic loading. The 3D frame discretization is performed by using two-node beam finite elements, with six degrees of freedom per node. Frame structures with shallow foundation, assumed to be rigid, can be modelled as rigidly connected to the soil at the soil surface, where the three components of the seismic motion are transmitted from the soil to frame base. The layered soil system is modelled as primary substructure and the framed structure as multi-connected secondary substructure, joined at the ground surface level.

The three-component (3C) seismic wave is propagated along a horizontally layered soil basin from soil-bedrock interface to the ground surface where a multistory multi-span frame structure is connected. The mutual influence of soil and structure on their response to seismic loading is studied. Modeling the three-component wave propagation, the mechanical coupling of multiaxial stress in the soil, inducing reduced soil strength and increasing nonlinear effects, can be analyzed. The incident direction of the seismic loading at the ground surface, shaking the frame structure base, can be taken into account. Different soil properties and soil stratigraphies, with consequent variation in nonlinear properties and impedance contrast between soil layers, modify seismic waves by amplification or deamplification effects.

The seismic response of soil profiles can be significantly different in the cases of free surface and the presence of a structure. The proposed model is a direct solution method, simultaneously modeling structure and ground motion by a global dynamic equilibrium equation for the soil-structure system. A one-directional threecomponent (1D-3C) propagation model (code SWAP_3C by Santisi d'Avila et al. [14], [15]) is adopted, where the three components of seismic waves are simultaneously propagated in one direction, from the soil-bedrock interface. An absorbing condition is assumed at soil-bedrock interface. A nonlinear constitutive relation of the Masing-Prandtl-Ishlinskii-Iwan type (MPII) is adopted for soil under multiaxial cyclic loading and a linear behavior is adopted for frame beams. The model is not dependent on the adopted constitutive relationships.

The three components of seismic motion are evaluated at the base of the 3D frame structure, taking into account soil nonlinearity, impedance contrast in a multilayered soil and soil-structure interaction. The ground motion is deduced considering the impedance contrast between soil and structure, due to their difference in terms of mass and stiffness.

The proposed 1D-3C propagation model taking into account SSI is implemented in a code called SFRINT_3C (Soil-FRame INTeraction_3Components).

A parametric analysis is undertaken to observe the effects of SSI for different combinations of soil features, dynamical properties of structures and earthquake frequency content. Ground motion time histories at the surface, profiles of stress, strain and motion components with depth and stress-strain hysteresis loops at a fixed depth are estimated for the soil stratification. Principal frequencies and modal shapes of the frame structure are evaluated, as well as deformation during the time history.

## 2 3-Component earthquake propagation in nonlinear soil

The three components of the seismic motion are propagated into a multilayered column of nonlinear soil from the top of the underlying elastic bedrock, by using a finite element scheme. Along the horizontal direction, at a given depth, the soil is assumed to be a continuous, homogeneous and infinite medium. Soil stratification is discretized into a system of horizontal layers, parallel to the $xy$ plane, using quadratic line elements with three nodes (Figure 1). There is not strain variation in $x$- and $y$-direction. Shear and pressure waves propagate vertically in $z$-direction.

### 2.1 Spatial discretization

The soil profile is discretized into $n_e$ quadratic line elements and consequently into $n_g = 2n_e + 1$ nodes (Figure 1), having as degrees of freedom the three displacements in directions $x$, $y$ and $z$. Finite element modeling of the horizontally layered soil system requires spatial discretization, to permit the problem solution, and the nonlinear mechanical behavior of soil demands time discretization of the process and linearization of the constitutive behavior in the time step. Accordingly, the incremental equilibrium equation in dynamic analysis, including compatibility conditions, three-dimensional constitutive relation and boundary conditions, is expressed in the matrix form as

$$\mathbf{M}_g \Delta \ddot{\mathbf{D}}_g + \mathbf{C}_g \Delta \dot{\mathbf{D}}_g + \mathbf{K}_g \Delta \mathbf{D}_g = \Delta \mathbf{F}_g \qquad (1)$$

where $\mathbf{D}_g$ is the assembled $3n_g$-dimensional nodal displacement vector of ground, $\dot{\mathbf{D}}_g$ and $\ddot{\mathbf{D}}_g$ are the velocity and acceleration vectors, respectively, i.e. the first and second time derivatives of the displacement vector. $\mathbf{M}_g$ and $\mathbf{K}_g$ are the assembled $(3n_g \times 3n_g)$-dimensional mass and stiffness matrix, respectively. $\mathbf{C}_g$ and $\mathbf{F}_g$ are the assembled $(3n_g \times 3n_g)$-dimensional damping matrix and the $3n_g$-dimensional load vector, respectively, derived from the imposed absorbing boundary condition, as explained in Section 2.2. The Finite Element Method, as applied in the present research, is completely described in the works of Batoz and Dhatt [3], Cook et al. [5] and Reddy [17].

The mass matrix $\mathbf{M}_g$ and stiffness matrix $\mathbf{K}_g$ result from the assemblage of $(9 \times 9)$-dimensional matrices as $\mathbf{M}^e$ and $\mathbf{K}^e$, respectively, corresponding to the element $e$, which are expressed by

$$\mathbf{M}^e = \rho_e A_e \int_0^{h_e} \mathbf{N}^T \mathbf{N}\, dz \qquad \mathbf{K}^e = A_e \int_0^{h_e} \mathbf{B}^T \mathbf{E} \mathbf{B}\, dz \qquad (2)\ a,b$$

where $h_e$ is the finite element length in vertical direction, $A_e$ is the element area in the horizontal plane and $\rho_e$ is the soil density assumed constant in the element.

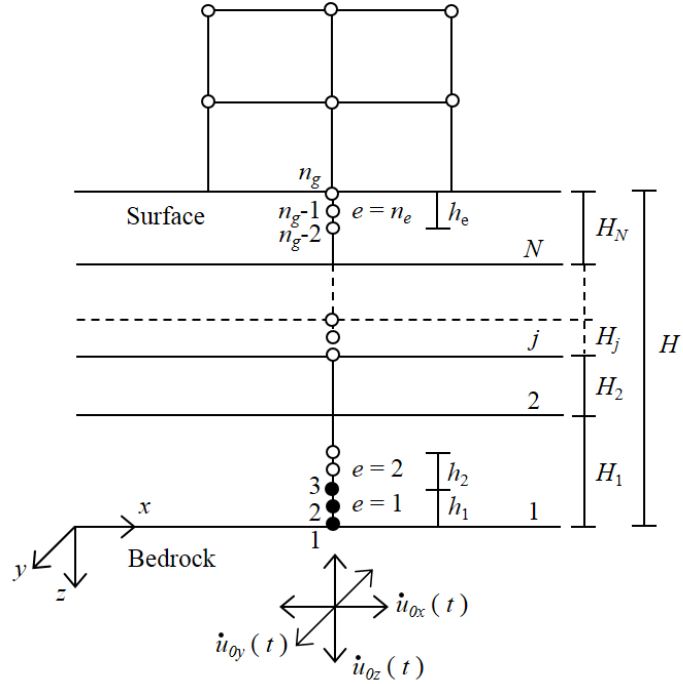

Figure 1: Spatial discretization of a horizontally layered soil, forced at its base by a 3-C earthquake and with a frame structure at the top.

The tangent constitutive $(6 \times 6)$-dimensional matrix $\mathbf{E}$, in equation (2)b, is evaluated by an incremental constitutive relationship as $\Delta \boldsymbol{\sigma}_g = \mathbf{E} \Delta \boldsymbol{\varepsilon}_g$, where the 6-dimensional stress and strain vectors are respectively defined as follows:

$$\boldsymbol{\sigma}_g = \begin{bmatrix} \sigma_{xx} & \sigma_{yy} & \tau_{xy} & \tau_{zy} & \tau_{zx} & \sigma_{zz} \end{bmatrix}^T$$
$$\boldsymbol{\varepsilon}_g = \begin{bmatrix} 0 & 0 & 0 & \gamma_{zy} & \gamma_{zx} & \varepsilon_{zz} \end{bmatrix}^T \qquad (3)\ a,b$$

according to the hypothesis of infinite horizontal soil. The wave propagation model is not dependent on the adopted constitutive relationship for soils.

In equation (2)a, $\mathbf{N}(z)$ is the $(3 \times 9)$-dimensional shape function matrix, such that $\mathbf{u}_g(z,t) = \mathbf{N}(\zeta(z))\mathbf{d}^e(t)$ (Cook et al. [5]), where the three terms of $\mathbf{u}_g$ are the soil displacements in $x$-, $y$- and $z$-direction and

$$\mathbf{d}^e = \begin{bmatrix} u_x^{2e-1} & u_y^{2e-1} & u_z^{2e-1} & \vdots & u_x^{2e} & u_y^{2e} & u_z^{2e} & \vdots & u_x^{2e+1} & u_y^{2e+1} & u_z^{2e+1} \end{bmatrix}^T \quad (4)$$

is the vector of displacements in directions $x$, $y$ and $z$ of the three nodes of element $e$. Quadratic shape functions, terms of the shape function matrix $\mathbf{N}$, corresponding to the three-node line element used to discretize the soil column, are defined according to Cook et al. [5].

The terms of the $(6 \times 9)$-dimensional matrix $\mathbf{B}(z)$, in equation (2)b, are the spatial derivatives of the shape functions, according to compatibility conditions and to the hypothesis of no strain variation in the horizontal directions $x$ and $y$. Given that the strain vector is related to displacement vector as $\boldsymbol{\varepsilon}_g = \partial \mathbf{u}_g$ and $\partial$ is a matrix of differential operators defined in such a way that compatibility equations are verified, consequently, it is $\boldsymbol{\varepsilon}_g = \mathbf{B}(\zeta(z))\mathbf{d}^e(t)$ and $\mathbf{B} = \partial \mathbf{N}$.

The damping matrix $\mathbf{C}^e$ and load vector $\mathbf{F}^e$ in equation (1) depend on boundary conditions and are defined in Section 2.2.

## 2.2 Boundary and initial conditions

The system of horizontal soil layers is bounded at the bottom by a semi-infinite elastic medium representing the seismic bedrock (Figure 2). The following condition, implemented by Joyner and Chen [11] in a finite difference formulation and used by Bardet and Tobita [2], is applied at the soil-bedrock interface to take into account the finite rigidity of the bedrock:

$$-\mathbf{p}^T \boldsymbol{\sigma} = \mathbf{c}(\dot{\mathbf{u}}_1 - 2\dot{\mathbf{u}}_0) \quad (5)$$

The stresses normal to the soil column base at the bedrock interface are $\mathbf{p}^T \boldsymbol{\sigma}$ and $\mathbf{c}$ is a $(3 \times 3)$-dimensional diagonal matrix whose terms are $\rho_b v_{sb}$, $\rho_b v_{sb}$ and $\rho_b v_{pb}$. The parameters $\rho_b$, $v_{sb}$ and $v_{pb}$ are density, shear and pressure wave velocities in the bedrock, respectively. According to equation (5), the damping matrix $\mathbf{C}^1$ and the load vector $\mathbf{F}^1$, for the first element $(e=1)$, are defined by

$$\mathbf{C}^1 = A_1 \left[ \mathbf{N}^T \mathbf{c} \mathbf{N} \right]_{z=0} \qquad \mathbf{F}^1 = A_1 \left[ \mathbf{N}^T \mathbf{c}(2\dot{\mathbf{u}}_0) \right]_{z=0} \quad (6)\text{ a,b}$$

$\mathbf{C}^e$ and $\mathbf{F}^e$ are a zero-matrix and vector, respectively, for the other elements all over the soil profile. The three terms of vector $\dot{\mathbf{u}}_1$ are the velocities at the soil-bedrock interface (node 1 in Figure 1) in $x$-, $y$- and $z$-direction, respectively. Terms of the 3-dimensional vector $\dot{\mathbf{u}}_0$ are the input incident velocities, in the underlying elastic medium, in directions $x$, $y$ and $z$, respectively. The 3-Component halved outcropping bedrock signals $\dot{\mathbf{u}}_0$ (Figure 2) are propagated.

The absorbing boundary condition (5), assumed at the soil-bedrock interface, allows energy to be radiated back into the underlying medium. This condition can be easily modified to use downhole records, assuming an imposed motion at the base of the soil profile (first node in Figure 1), according to Santisi d'Avila and Semblat [16].
The soil profile is bounded at the bottom by the elastic bedrock and it is connected to the frame structure at the top. Global equilibrium equation for the soil-frame system is directly solved, by imposing boundary conditions only at the soil-bedrock interface.
Gravity load is imposed as static initial condition in terms of strain and stress in each node.

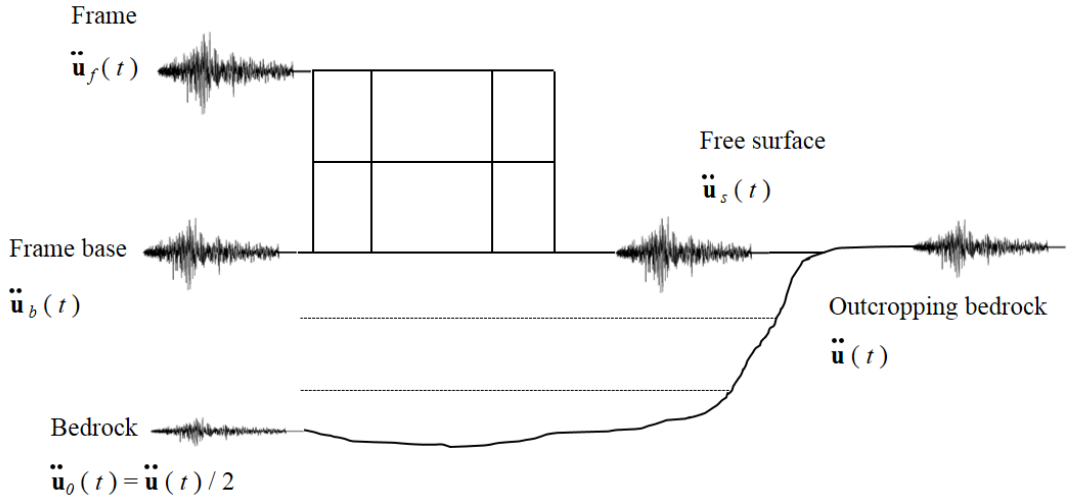

Figure 2: Representation of seismic signals at different points of the model scheme: outcropping bedrock, base of the soil profile, free surface at the top of soil profile, base and top of the frame structure.

## 2.3 Constitutive model

Modeling the propagation of a three-component earthquake in stratified soils requires a three-dimensional constitutive model for soil. The so-called Masing-Prandtl-Ishlinskii-Iwan (MPII) constitutive model, suggested by Iwan [9] and applied by Joyner [10] and Joyner and Chen [11] in a finite difference formulation, is used in the present work to properly model the nonlinear soil behavior in a finite element scheme. The MPII model is used to represent the behavior of materials satisfying Masing criterion [12] and not depending on the number of loading cycles. The stress level depends on the strain increment and strain history but not on the strain rate. This rheological model has no viscous damping. The energy dissipation process is purely hysteretic and does not depend on the frequency. Iwan [9] proposes an extension of the standard incremental theory of plasticity (Fung [6]), modifying the 1D approach by introducing a family of yield surfaces. He models nonlinear stress-strain curves using a series of mechanical elements, having different stiffness and increasing sliding resistance. The MPII model takes into account the nonlinear hysteretic behavior of soils in a three-dimensional stress state, using an elasto-plastic approach with hardening, based on the definition of a series of nested yield surfaces, according to von Mises' criterion. The shear modulus is strain-dependent. The MPII hysteretic model for dry soils, used in the present research, is applied for strains in the range of stable nonlinearity.

The main feature of the MPII rheological model is that the only necessary input data, to identify soil properties in the applied constitutive model, is the shear modulus decay curve $G(\gamma)$ versus shear strain $\gamma$. The initial elastic shear modulus $G_0 = \rho v_s^2$, depends on the mass density $\rho$ and the shear wave velocity in the medium $v_s$. The P-wave modulus $M = \rho v_p^2$, depending on the pressure wave velocity in the medium $v_p$, characterizes the longitudinal behavior of soil.

In the present study the soil behavior is assumed adequately described by a hyperbolic stress-strain curve (Hardin and Drnevich [7]). This assumption yields a normalized shear modulus decay curve, used as input curve representing soil characteristics, expressed as

$$G/G_0 = 1/(1 + |\gamma/\gamma_r|) \tag{7}$$

where $\gamma_r$ is a reference shear strain provided by test data corresponding to an actual tangent shear modulus equivalent to $50\%$ of the initial shear modulus. The applied constitutive model (Iwan [9]; Joyner 1975 [10]; Joyner and Chen 1975 [11]) does not depend on the hyperbolic backbone curve. It could incorporate also shear modulus decay curves obtained from laboratory dynamic tests on soil samples.

According to Joyner [10], the actual strain level and the strain and stress values at the previous time step allow to evaluate the tangent constitutive matrix $\mathbf{E}$ in equation (2) and the stress increment $\Delta \boldsymbol{\sigma}_g$ (Santisi d'Avila et al. [14]).

## 3 3D frame structure modeling under 3C seismic loading

The response of a regular frame composed by horizontal and vertical beam elements, along three orthogonal directions $x$, $y$ and $z$, shaken by the three components of a seismic motion, is modelled. Beams are assumed composed by a continuous and homogeneous medium with constant cross-section along their longitudinal axis. The hypothesis of plane cross-section, not necessarily perpendicular to the beam axis, is assumed for beams during deformation. Beam cross-sectional parameters are the constant area $A$, the moments of inertia $I_y$ and $I_z$ with respect to $y$ and $z$ axis, respectively, shape factors $\chi_y$ and $\chi_z$ for transverse shear and the second moment of area $J$. Material parameters are the compression modulus $E$, shear modulus $G = E/(2(1+\nu))$, where $\nu$ is the Poisson's ratio, and mass density $\rho$.

## 3.1 Spatial discretization

The 3D frame structure is modelled by a system of one-dimensional 2-node beam elements. Each node has 6 degrees of freedom in the $xyz$ global coordinate system, that are the displacements in *x*-, *y*- and *z*-direction and rotations around the same axes. The 6-dimensional displacement vector of a generic node in a beam *e*, parallel to *x* -axis, is defined as

$$\mathbf{u}_f = \begin{bmatrix} u_x & u_y & u_z & \theta_x & \theta_y & \theta_z \end{bmatrix}^T \quad (8)$$

The corresponding 6-dimensional vector of nontrivial strains for a 3D beam is

$$\boldsymbol{\varepsilon}_f = \begin{bmatrix} \varepsilon_{xx} & 0 & 0 & 0 & \gamma_{xz} & \gamma_{xy} \end{bmatrix}^T \quad (9)$$

The incremental form of dynamic equilibrium equation of the analyzed 3D frame in matrix form is

$$\mathbf{M}_f \Delta \ddot{\mathbf{D}}_f + \mathbf{C}_f \Delta \dot{\mathbf{D}}_f + \mathbf{K}_f \Delta \mathbf{D}_f = \mathbf{0} \quad (10)$$

The dimension of equation (10) is $6n_{fb}$, where $n_{fb}$ is the number of nodes of the frame including the base. $\mathbf{D}_f$, $\dot{\mathbf{D}}_f$ and $\ddot{\mathbf{D}}_f$ are the $6n_{fb}$-dimensional assembled vector of nodal displacement, velocity and acceleration, respectively. The consistent mass matrix $\mathbf{M}_f$, the damping matrix $\mathbf{C}_f$ and stiffness matrix $\mathbf{K}_f$ are assembled (Reddy [17]) according to the geometry of the frame. The $(6n_{fb} \times 6n_{fb})$-dimensional consistent mass matrix $\mathbf{M}_f$ and stiffness matrix $\mathbf{K}_f$ result from the assemblage of $(12 \times 12)$-dimensional matrices as $\mathbf{M}^e$ and $\mathbf{K}^e$, respectively, of each beam element *e*, in global coordinates $xyz$, which are expressed as (Batoz and Dhatt [3])

$$\mathbf{M}^e = \boldsymbol{\Lambda}_e^T \overline{\mathbf{M}}^e \boldsymbol{\Lambda}_e \qquad \mathbf{K}^e = \boldsymbol{\Lambda}_e^T \overline{\mathbf{K}}^e \boldsymbol{\Lambda}_e \quad (11) \text{ a,b}$$

where $\boldsymbol{\Lambda}_e^T$ is the $(12 \times 12)$-dimensional rotation matrix, associated to beam *e*, that allow to transform displacements and rotations from local coordinates $\overline{xyz}$ to global coordinates $xyz$. $\overline{\mathbf{M}}^e$ and $\overline{\mathbf{K}}^e$, expressed in local coordinates $\overline{xyz}$, for a 2-node beam element *e* of length $L_e$ in $\overline{x}$ direction, are

$$\overline{\mathbf{M}}^e = \rho A \int_0^{L_e} \mathbf{N}^T \mathbf{N} \, d\overline{x} \qquad \overline{\mathbf{K}}^e = \int_0^{L_e} \mathbf{B}^T \mathbf{E} \mathbf{B} \, d\overline{x} \quad (12) \text{ a,b}$$

where $A$ is the beam cross-sectional area and $\rho$ is the material density assumed constant in the element.

A 2-node interdependent interpolation element (Reddy [17]) is used in a finite element scheme, based on Hermite cubic interpolation of displacements and an interdependent quadratic interpolation of rotations, so that displacement first derivative is a polynomial with the same degree of rotations.

The $(6 \times 12)$-dimensional shape function matrix $\mathbf{N}$ is defined according to the transformation $\mathbf{u}_f(x,t) = \mathbf{N}(\xi(x))\mathbf{d}^e(t)$, for each beam element $e$ having nodes $j$ and $l$, where

$$\mathbf{d}^e = \begin{bmatrix} \mathbf{d}_j & \mathbf{d}_l \end{bmatrix}^T = \begin{bmatrix} u_x^j & u_y^j & u_z^j & \theta_x^j & \theta_y^j & \theta_z^j & u_x^l & u_y^l & u_z^l & \theta_x^l & \theta_y^l & \theta_z^l \end{bmatrix}^T \quad (13)$$

is a 12-dimensional time dependent node displacement vector. Shape functions in matrix $\mathbf{N}$ are defined according to Reddy [17]. The $(6 \times 12)$-dimensional matrix $\mathbf{B}$, in equation (12)b, is defined in such a way that compatibility equations $\boldsymbol{\varepsilon}_f = \partial \mathbf{u}_f$ are verified, where $\partial$ is a matrix of differential operators.

The $(6 \times 6)$-dimensional constitutive matrix $\mathbf{E}$, in equation (12)b, is defined according with an incremental constitutive relationship such as $\Delta \mathbf{S} = \mathbf{E} \Delta \boldsymbol{\varepsilon}_f$, where

$$\Delta \mathbf{S} = \begin{bmatrix} \Delta N & \Delta V_y & \Delta V_z & \Delta M_x & \Delta M_y & \Delta M_z \end{bmatrix}^T \quad (14)$$

is the vector of internal forces. The frame model is independent from the selected constitutive law. In this research, a linear behavior for the frame structure material is assumed, with $\mathbf{E} = \mathrm{diag}(EA, \chi_y GA, \chi_z GA, GJ, EI_y, EI_z)$.

Damping matrix $\mathbf{C}_f$ in equation (10) depend on modal analysis and it is defined in Section 3.3.

## 3.2 Boundary and initial conditions

The 3D frame is rigidly connected at the ground surface. Consequently, rotations of the frame base nodes are assumed null.
The shallow foundation is assumed to be rigid. Accordingly, all nodes of the frame base are supposed to be submitted to the same ground motion.
Static loading configuration represents the initial condition for the frame structure. The linear elastic solution of static equilibrium equation of the analyzed 3D frame is

$$\mathbf{K}_f \mathbf{D}_f = \mathbf{F}_f - \mathbf{R} \quad (15)$$

where $\mathbf{F}_f$ is a $6n_{fb}$-dimensional vector of static nodal loads, obtained by assembling the 6-dimensional load vectors of frame nodes

$$\mathbf{f} = \begin{bmatrix} F_x & F_y & F_z & M_x & M_y & M_z \end{bmatrix}^T \qquad (16)$$

which terms are the external forces directly applied in each node. The $6n_{fb}$-dimensional vector $\mathbf{R}$ is assembled by using the $12$-dimensional reaction force vectors that are $\mathbf{R}^e = \mathbf{\Lambda}_e^T \overline{\mathbf{R}}^e$, for each beam $e$, in global coordinate system. The terms of vector $\overline{\mathbf{R}}^e$ are reaction forces to loads in beam $e$, expressed in local coordinates $\overline{xyz}$.

## 3.3 Modal analysis

The system of equations (10) is composed by $6n_b$ equations related to the motion of frame base nodes and $6n_f$ equations corresponding to the motion of the other frame nodes. Accordingly, equation (10) can be written as

$$\begin{bmatrix} \mathbf{M}_{bb}^f & \mathbf{M}_{bf} \\ \mathbf{M}_{fb} & \mathbf{M}_{ff} \end{bmatrix} \begin{bmatrix} \Delta \ddot{\mathbf{D}}_b \\ \Delta \ddot{\mathbf{D}}_{ff}^t \end{bmatrix} + \begin{bmatrix} \mathbf{C}_{bb}^f & \mathbf{C}_{bf} \\ \mathbf{C}_{fb} & \mathbf{C}_{ff} \end{bmatrix} \begin{bmatrix} \Delta \dot{\mathbf{D}}_b \\ \Delta \dot{\mathbf{D}}_{ff}^t \end{bmatrix} + \begin{bmatrix} \mathbf{K}_{bb}^f & \mathbf{K}_{bf} \\ \mathbf{K}_{fb} & \mathbf{K}_{ff} \end{bmatrix} \begin{bmatrix} \Delta \mathbf{D}_b \\ \Delta \mathbf{D}_{ff}^t \end{bmatrix} = \begin{bmatrix} \mathbf{0} \\ \mathbf{0} \end{bmatrix} \qquad (17)$$

where $f$ and $b$ indicate each term related with frame and boundary (soil-frame interface), respectively.

Fundamental fixed-base frequencies of the frame structure are obtained solving the $6n_f$-dimensional eigenproblem $(\mathbf{K}_{ff} - \omega^2 \mathbf{M}_{ff})\mathbf{\Phi} = \mathbf{0}$, where $\omega$ are the natural angular frequencies and $\mathbf{\Phi}$ is the modal matrix. The mass-normalization of matrix $\mathbf{\Phi}$ implies $\mathbf{\Phi}^T \mathbf{M}_{ff} \mathbf{\Phi} = \mathbf{I}$ and $\mathbf{\Phi}^T \mathbf{K}_{ff} \mathbf{\Phi} = \mathbf{\Omega}^2$, where $\mathbf{\Omega} = \text{diag}(\omega_1, \dots \omega_{n_f})$ and $\mathbf{I}$ is a $6n_f$-dimensional identity matrix.

Damping matrix can be evaluated as $\mathbf{C}_{ff} = \mathbf{\Phi}^{T-1} \mathbf{\Xi} \mathbf{\Phi}^{-1} = \mathbf{M}_{ff} \mathbf{\Phi} \mathbf{\Xi} \mathbf{\Phi}^T \mathbf{M}_{ff}$ (Chopra [4]), where $\mathbf{\Xi} = 2\zeta_0 \mathbf{\Omega} = \text{diag}(2\zeta_0 \omega_1, \dots, 2\zeta_0 \omega_{n_f})$. The damping ratio $\zeta_0$ is know for typical materials employed in regular 3D frames. It is assumed $\zeta_0 = 0.05$ for typical reinforced concrete buildings. Otherwise, damping matrix can be estimated by the Rayleigh approach, as $\mathbf{C}_{ff} = a_0 \mathbf{M}_{ff} + a_1 \mathbf{K}_{ff}$, depending on mass and stiffness matrices, with $a_0 = 2\zeta_0 \omega_1 \omega_2 / (\omega_1 + \omega_2)$ and $a_1 = 2\zeta_0 / (\omega_1 + \omega_2)$. $\mathbf{C}_{bb}^f$, $\mathbf{C}_{bf}$ and $\mathbf{C}_{fb}$ are null.

## 4 Soil-structure interaction modeling

The 3-Component motion at the base of frame structure is assumed coincident with ground motion at the top of soil profile (Figure 3). The same loading motion is applied to all column bases, reducing degrees of freedom of the frame base to only three displacements at the soil-frame interface level ($6n_b = 3$). Rigid rotations of the foundation are assumed null, supposing that surface waves are negligible, according to the employed 1D wave propagation model. This analysis considers inertial interaction modifying the ground motion, due to inertia efforts induced by structure mass at the soil-structure connection level. Kinematic interaction, induced by stiffness variation between soil and structure foundation at the most surface soil layers, is assumed negligible for shallow foundations and vertical propagation.

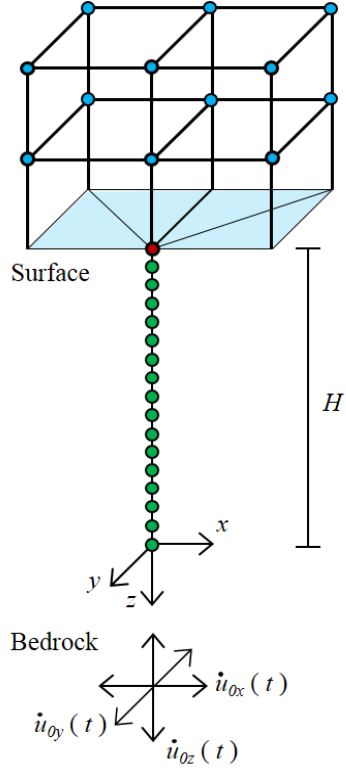

Figure 3: Spatial discretization of soil (3-node line elements) and frame (2-node beam elements) with a connection node and elastic boundary condition.

The equilibrium condition under dynamic loading for the soil-frame system is defined according to the concept of coupling a primary substructure with a multi-connected secondary substructure, joined at the ground surface level (Figure 1). Equations (1) can be written as

$$\begin{bmatrix} \mathbf{M}_{gg} & \mathbf{M}_{gb} \\ \mathbf{M}_{bg} & \mathbf{M}_{bb}^g \end{bmatrix} \begin{bmatrix} \Delta \ddot{\mathbf{D}}_{gg} \\ \Delta \ddot{\mathbf{D}}_b \end{bmatrix} + \begin{bmatrix} \mathbf{C}_{gg} & \mathbf{C}_{gb} \\ \mathbf{C}_{bg} & \mathbf{C}_{bb}^g \end{bmatrix} \begin{bmatrix} \Delta \dot{\mathbf{D}}_{gg} \\ \Delta \dot{\mathbf{D}}_b \end{bmatrix} + \begin{bmatrix} \mathbf{K}_{gg} & \mathbf{K}_{gb} \\ \mathbf{K}_{bg} & \mathbf{K}_{bb}^g \end{bmatrix} \begin{bmatrix} \Delta \mathbf{D}_{gg} \\ \Delta \mathbf{D}_b \end{bmatrix} = \begin{bmatrix} \Delta \mathbf{f}_g \\ \mathbf{0} \end{bmatrix} \quad (18)$$

indicating with $g$ and $b$ each term related with ground and boundary (interface between soil and frame), respectively. $\mathbf{C}_{bb}^g$, $\mathbf{C}_{bg}$ and $\mathbf{C}_{gb}$ are null. The only non-zero terms in $\mathbf{C}_{gg}$ and $\Delta \mathbf{f}_g$ are those related to the node at soil-bedrock interface, according to the adopted boundary condition (see Section 2.2). The total dimension of equation system (18) is $3n_g = 6n_e + 3$. It is decomposed in two blocks of $6n_e$ equations related to soil motion variables, with $n_e$ the number of finite elements employed in the soil discretization, and 3 equations associated to ground motion at the surface.

Equation (17) is rewritten as

$$\begin{bmatrix} \mathbf{M}_{bb}^f & \mathbf{M}_{bf} \\ \mathbf{M}_{fb} & \mathbf{M}_{ff} \end{bmatrix} \begin{bmatrix} \Delta \ddot{\mathbf{D}}_b \\ \Delta \ddot{\mathbf{D}}_{ff}^t \end{bmatrix} + \begin{bmatrix} 0 & 0 \\ 0 & \mathbf{C}_{ff} \end{bmatrix} \begin{bmatrix} \Delta \dot{\mathbf{D}}_b \\ \Delta \dot{\mathbf{D}}_{ff}^t \end{bmatrix} + \begin{bmatrix} \mathbf{K}_{bb}^f & \mathbf{K}_{bf} \\ \mathbf{K}_{fb} & \mathbf{K}_{ff} \end{bmatrix} \begin{bmatrix} \Delta \mathbf{D}_b \\ \Delta \mathbf{D}_{ff}^t \end{bmatrix} = \begin{bmatrix} 0 \\ 0 \end{bmatrix} \quad (19)$$

considering that $\mathbf{C}_{bb}^f$, $\mathbf{C}_{bf}$ and $\mathbf{C}_{fb}$ are null (see Section 3.3).

The total dimension of equation system (19) is $6n_{fb} = 6n_b + 6n_f = 3 + 6n_f$. It is composed of two blocks of 3 equations related to the motion of frame base, according to the hypothesis of equal seismic loading at the base of all frame columns and negligible base rotation, and $6n_f$ equations corresponding to frame node motion. A condensation of $6n_b$ degrees of freedom of base nodes is applied, allowing to consider only 3 degrees of freedom.

Combining equations (18) and (19), we obtain the following equilibrium equation:

$$\mathbf{M}\,\Delta\ddot{\mathbf{D}} + \mathbf{C}\,\Delta\dot{\mathbf{D}} + \mathbf{K}\,\Delta\mathbf{D} = \Delta\mathbf{F} \quad (20)$$

where the $\big((6n_g + 3 + 6n_f) \times (6n_g + 3 + 6n_f)\big)$-dimensional mass, damping and stiffness matrices are, respectively,

$$\mathbf{M} = \begin{bmatrix} \mathbf{M}_{gg} & \mathbf{M}_{gb} & 0 \\ \mathbf{M}_{bg} & \mathbf{M}_{bb}^g + \mathbf{M}_{bb}^f & \mathbf{M}_{bf} \\ 0 & \mathbf{M}_{fb} & \mathbf{M}_{ff} \end{bmatrix} \quad \mathbf{C} = \begin{bmatrix} \mathbf{C}_{gg} & 0 & 0 \\ 0 & 0 & 0 \\ 0 & 0 & \mathbf{C}_{ff} \end{bmatrix} \quad \mathbf{K} = \begin{bmatrix} \mathbf{K}_{gg} & \mathbf{K}_{gb} & 0 \\ \mathbf{K}_{bg} & \mathbf{K}_{bb}^g + \mathbf{K}_{bb}^f & \mathbf{K}_{bf} \\ 0 & \mathbf{K}_{fb} & \mathbf{K}_{ff} \end{bmatrix} \quad (21)$$

The load and displacement increment vectors are, respectively,

$$\Delta\mathbf{F} = \begin{bmatrix} \Delta\mathbf{f}_g & 0 & 0 \end{bmatrix}^T$$
$$\Delta\mathbf{D} = \begin{bmatrix} \Delta\mathbf{D}_{gg} & \Delta\mathbf{D}_b & \Delta\mathbf{D}_{ff}^t \end{bmatrix}^T \quad (22)\text{a,b}$$

The $6n_f$-dimensional relative displacement vector of frame nodes is evaluated as $\Delta\mathbf{D}_{ff} = \Delta\mathbf{D}_{ff}^t - \Delta\mathbf{D}_{ff}^s$, where $\Delta\mathbf{D}_{ff}^t$ is the vector of total displacement increment and $\Delta\mathbf{D}_{ff}^s = -\mathbf{K}_{ff}^{-1}\mathbf{K}_{fb}\Delta\mathbf{D}_b$ is the displacement vector due to the static application of base node displacement $\Delta\mathbf{D}_b$ (Chopra [4]).

## 4.1 Time discretization

Time integration is done according to Newmark's process. The incremental dynamic equilibrium equation (20) can be written as

$$\mathbf{M}_k \Delta \ddot{\mathbf{D}}_k^i + \mathbf{C}_k^i \Delta \dot{\mathbf{D}}_k^i + \mathbf{K}_k^i \Delta \mathbf{D}_k^i = \Delta \mathbf{F}_k \quad (23)$$

according to time discretization. The subscript $k$ indicates the time step $t_k$ and $i$ the iteration of the problem solving process. Equation (23) becomes

$$\tilde{\mathbf{K}}_k^i \Delta \mathbf{D}_k^i = \Delta \mathbf{F}_k + \mathbf{A}_k^i \quad (24)$$

The equivalent stiffness matrix and the equivalent load vector are, respectively,

$$\tilde{\mathbf{K}}_k^i = 1/(\beta \Delta t^2) \mathbf{M}_k + \alpha/(\beta \Delta t) \mathbf{C}_k^i + \mathbf{K}_k^i$$

$$\mathbf{A}_k^i = \left[ 1/(\beta \Delta t) \mathbf{M}_k + \alpha/\beta \, \mathbf{C}_k^i \right] \dot{\mathbf{D}}_{k-1} + \left[ 1/(2\beta) \mathbf{M}_k + (\alpha/(2\beta) - 1) \Delta t \, \mathbf{C}_k^i \right] \ddot{\mathbf{D}}_{k-1} \quad (25)\text{a,b}$$

Equation (24) requires an iterative solving, at each time step $k$, to correct the tangent stiffness matrix $\mathbf{K}_k^i$. Starting from the stiffness matrix $\mathbf{K}_k^1 = \mathbf{K}_{k-1}$, evaluated at the previous time step, the value of matrix $\mathbf{K}_k^i$ is updated at each iteration $i$. An elastic linear behavior is assumed for the first iteration at the first time step. The nodal displacement $\mathbf{D}_k^i = \mathbf{D}_{k-1} + \Delta \mathbf{D}_k^i$ is obtained and strain increments are deduced from the displacement increments $\Delta \mathbf{D}_k^i = \mathbf{D}_k^i - \mathbf{D}_{k-1}$. Stress increments and the tangent constitutive matrix $\mathbf{E}_k^i$ are obtained through the constitutive relationships for the soil. Matrices $\mathbf{K}_k^i$, $\mathbf{C}_k^i$ and the equivalent stiffness matrix $\tilde{\mathbf{K}}_k^i$ are then calculated and the process restarts. The correction process continues until the difference between two successive approximations is reduced to a fixed tolerance, according to $\left| \mathbf{D}_k^i - \mathbf{D}_k^{i-1} \right| < \eta \left| \mathbf{D}_k^i \right|$, where $\eta = 10^{-3}$.

Total nodal velocity and acceleration are evaluated by

$$\begin{cases} \dot{\mathbf{D}}_k^i = \dot{\mathbf{D}}_{k-1} + \alpha/(\beta \Delta t) \Delta \mathbf{D}_k^i - \alpha/\beta \, \dot{\mathbf{D}}_{k-1} + (1 - \alpha/(2\beta)) \Delta t \, \ddot{\mathbf{D}}_{k-1} \\ \ddot{\mathbf{D}}_k^i = \ddot{\mathbf{D}}_{k-1} + 1/(\beta \Delta t^2) \Delta \mathbf{D}_k^i - 1/(\beta \Delta t) \dot{\mathbf{D}}_{k-1} - 1/(2\beta) \ddot{\mathbf{D}}_{k-1} \end{cases} \quad (26)$$

Afterwards, the next time step is analyzed. The hypothesis of linear acceleration in the time step is assumed and the choice of the two parameters $\beta = 0.3025$ and $\alpha = 0.6$ guarantees unconditional stability of the time integration scheme and numerical damping properties to damp higher modes (Hughes [8]).

# 5 Analysis of the local soil-structure interaction

The influence of 3-Component shaking vs 1C motion and local effects dues to impedance contrast in multilayered soil are extensively described by Santisi et al. [14], where the same 1D-3C wave propagation model is used. In this research, coupling of soil and frame is investigated, to show the interaction effects reproduced by a one-directional wave propagation model assembled with 3D multi-story multi-span frame model. Concerning the seismic loading, the acceleration signal adopted in this analysis is a record from the 2009 L'Aquila earthquake in Central Italy. The halved acceleration (Figure 4), having peak ground acceleration (PGA) equal to 1.14 m/s$^2$, is integrated and forced at the base of soil profile.

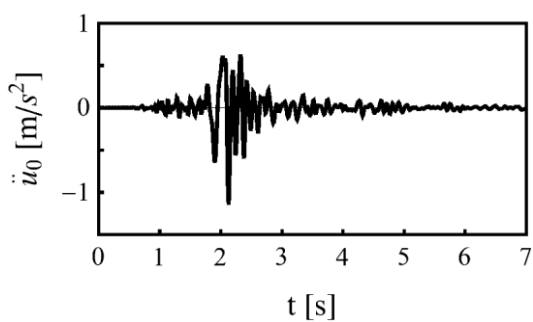

Figure 4: One-component acceleration record from the 2009 L'Aquila earthquake.

The proposed model is verified by comparison with GEFDyn code (Aubry et al. [1]), in the case of 1C-propagation, linear behavior of soil and structure having one degree of freedom.

## 5.1 SSI vs free field condition

A one-story one-span frame on the homogeneous soil profile S1 is analyzed. Stratigraphy and soil properties, as the density $\rho$ and shear and pressure velocity in the medium, $v_s$ and $v_p$, respectively, are reported in Table 1. Beam element dimensions, vertical load $g$, damping ratio $\zeta_0$ and the adopted material properties (compression modulus E, Poisson ratio $v$ and density $\rho$) are described in Table 2 (frame F).
The influence of soil column cross-sectional area in SSI effect is investigated, by evaluating the soil-surface-bedrock transfer function (TF), the top-frame-soil-surface TF and acceleration, velocity, shear stress and strain profiles with depth, for different values of the side $a$ of soil square cross-section, assumed constant with depth. The principal frequency of soil profile $v_s/(4H) = 3.75\,\text{Hz}$ is reproduced in all the cases, with $v_s$ and $H$ the shear velocity in the medium and the soil column height, respectively (Figure 5a). When adopting $a = 1\,\text{km}$, the SSI is not observed. The soil mass is predominant and a solution close to the free-field condition is obtained (FF curve in Figure 5). On the contrary, when adopting $a \leq 30\,\text{m}$, the effect of the structure is overestimated. Shear strains and stresses appear not null at the soil-frame interface (Figure 5c), that represents a not realistic solution. The fixed base behavior is well reproduced for $a \geq 50\,\text{m}$. A soil column with side $a = 50\,\text{m}$ is adopted in the following computations.

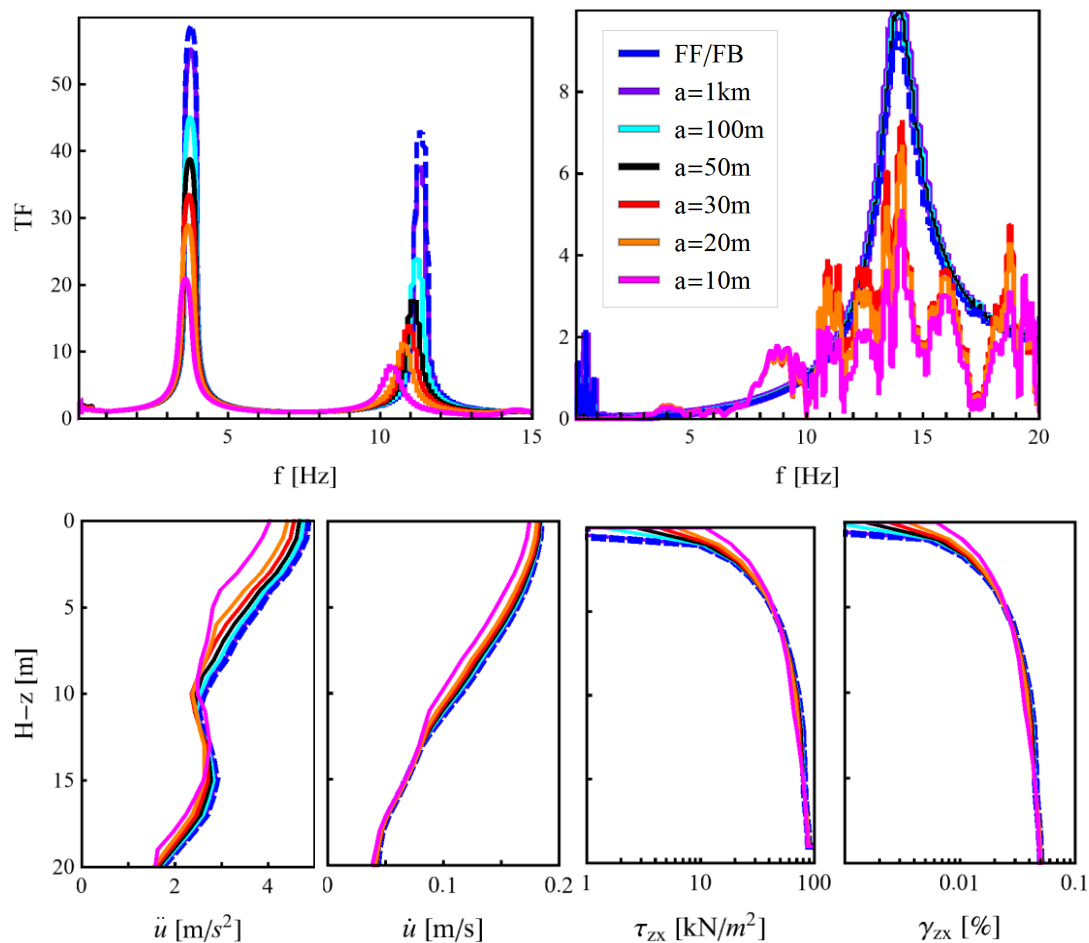

Figure 5: (a) Soil-surface-bedrock Transfer Function. (b) Top-frame-soil-surface Transfer Function. (c) Acceleration, velocity, stress and strain profiles with depth.

| z | H | $\rho$ | $v_p$ | $v_s$ | z | H | $\rho$ | $v_p$ | $v_s$ |
| m | m | kg/m$^3$ | m/s | m/s | m | m | kg/m$^3$ | m/s | m/s |
| --- | --- | --- | --- | --- | --- | --- | --- | --- | --- |
| 0 - 20 | 20 | 2000 | 700 | 300 | 0 - 20 | 20 | 2000 | 1150 | 500 |
| > 20 | | 2500 | 1900 | 1000 | > 20 | | 2500 | 1900 | 1000 |

Table 1: Properties of the homogenous soil profiles S1 (left) and R1 (right).

Using a 1D soil model, the soil perceives the building as an elementary oscillator (one degree of freedom) characterized by its mass and stiffness. The number of storeys and spans modifies the total mass and stiffness and, consequently, the SSI effect, but the influence of an increasing floor area, due to an increasing number of spans, is not captured by a 1D soil model. For this reason, the same soil area is assumed in all cases in this analysis.

| Frame | Floors | Spans x | Spans y | Section cm | L m | E N/mm$^2$ | $\nu$ | $\rho$ kg/m$^3$ | g kN/m | $\zeta_0$ % |
|---|---|---|---|---|---|---|---|---|---|---|
| F | 1 | 1 | 1 | 30x60 | 3 | 31220 | 0.2 | 2500 | 13 | 5 |
| R | 3 | 1 | 1 | 40x90 | 3 | 31220 | 0.2 | 2500 | 13 | 5 |
| S | 3 | 1 | 1 | 30x50 | 3 | 31220 | 0.2 | 2500 | 13 | 5 |
| T | 3 | 1 | 1 | 30x90 | 3 | 31220 | 0.2 | 2500 | 13 | 5 |

Table 2: Properties of analyzed frame structures.

## 5.2 Influence of frequency content

The single-frequency Mavroeidis-Papageorgiou wavelet is adopted to study the effect of frequency content of ground motion, shaking rigid and soft soil profiles, coupled with rigid and soft frame structures at the top. The outcrop motion is obtained by the following expression:

$$\ddot{u}_0(t) = \ddot{u}_{0\max}/2 \left[1+\cos\left(2\pi f/n\left(t-t_f/2\right)\right)\right]\cos\left(2\pi f\left(t-t_f/2\right)\right) \quad 0 < t \leq 2t_f \quad (27)$$

where $\ddot{u}_{0\max}$, $f$ and $t_f = n/(2f)$ are the acceleration peak, principal frequency and duration, respectively. The number of peaks is assumed $n = 5$ and $\ddot{u}_{0\max} = 3.5\,\text{m}/s^2$. Acceleration signals are halved to remove the free surface effect and integrated, to obtain incident velocities, before being forced at the base of the soil profile.

The influence of earthquake frequency content in structural response is analyzed studying the behavior of a three-story one-span rigid frame R (principal frequency $f = 5.8\,\text{Hz}$) shaked by an 1C incident wave having frequency $f_0 = 6\,\text{Hz}$, in both cases of rigid soil R1 ($f_g = 6.25\,\text{Hz}$) and softer soil S1 ($f_g = 3.75\,\text{Hz}$). This situation is compared with the behavior of a softer frame S (principal frequency $f = 3.5\,\text{Hz}$) shaked by an incident wave having frequency $f_0 = 3.5\,\text{Hz}$, in both cases of rigid soil R1 and softer soil S1. Beam element dimensions, vertical load $g$, damping ratio $\zeta_0$ and the adopted material properties are described in Table 2 for the analyzed rigid and softer frame.

This analysis confirms that a rigid frame in a soft soil is less stressed by a seismic loading having a frequency content close to the frame principal frequency, than in the case of rigid soil (having principal frequency close to that of frame and quake). Similarly, a soft frame in a rigid soil is less stressed by a quake having a frequency content close to the its principal frequency, than in the case of softer soil (having principal frequency close to that of frame and quake). Numerical results in terms of horizontal acceleration at the top of the frame structure are showed in Table 3.

|       | f (Frame) | $f_g$ (Soil) | $f_0$ (Quake) | ü      |
|-------|-----------|--------------|---------------|--------|
|       | Hz        | Hz           | Hz            | m/s$^2$|
| R-R1  | 5.8       | 6.25         | 6.0           | 8.90   |
| R-S1  | 5.8       | 3.75         | 6.0           | 4.69   |
| S-S1  | 3.5       | 3.75         | 3.5           | 20.91  |
| S-R1  | 3.5       | 6.25         | 3.5           | 12.14  |

Table 3: Horizontal acceleration at the top of the frame structure.

## 5.3 Influence of soil nonlinearity

Nonlinear features are characterized, in the adopted MPII constitutive model for soils, by the shear modulus decay curve. A hyperbolic first loading curve is assumed and the shear modulus decay curve is defined by Equation (7). The shear strain $\gamma_r$, related with a 50% decay of shear modulus, is varied to increase nonlinearity in the multilayered soil profile (Table 4) and observe the associated effects. The applied incident wave is the 1C signal shown in Figure 4.

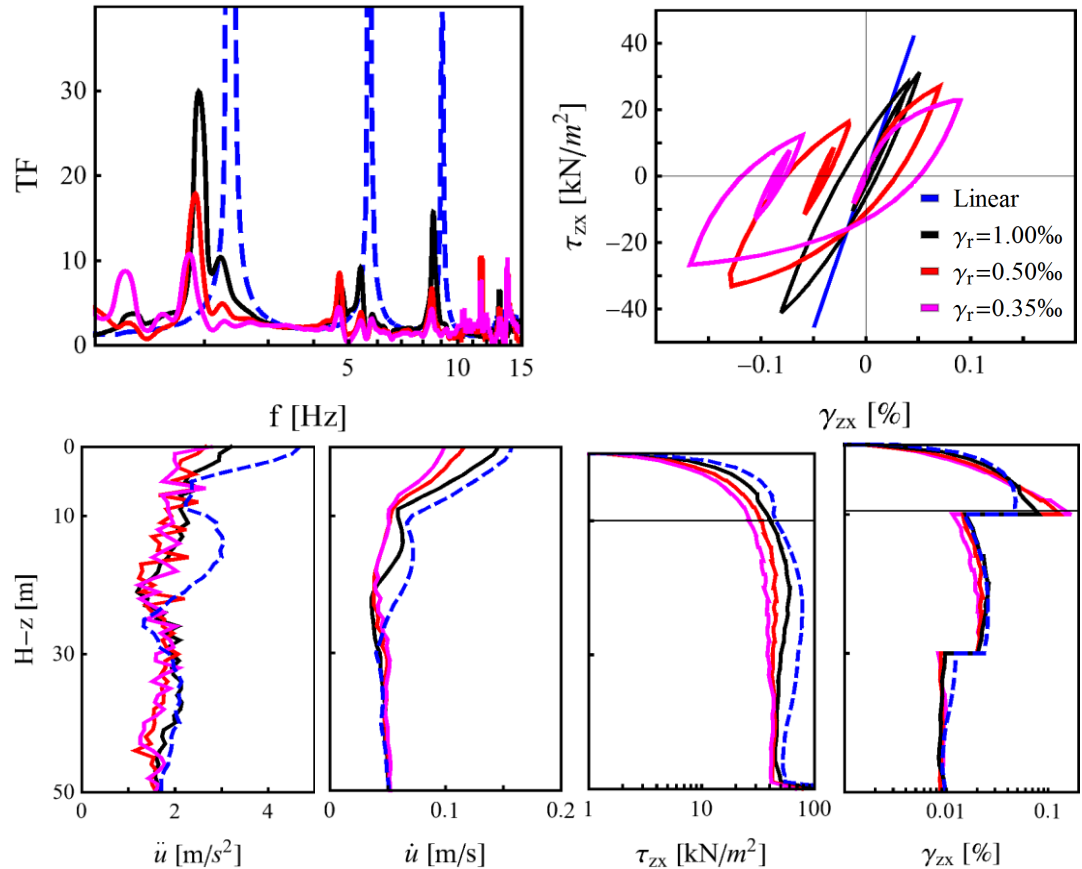

Figure 6: (a) Soil-surface-bedrock Transfer Function. (b) Shear stress-shear strain hysteresis loop. (c) Acceleration, velocity, stress and shear strain profiles with depth.

Nonlinear effects lead to strength reduction, strain increasing, and decrement of principal frequencies (Figure 6). Nonlinear behavior can reduce ground motion peaks at the ground surface.

| z | H | ρ | $v_p$ | $v_s$ |
| --- | --- | --- | --- | --- |
| m | m | kg/m$^3$ | m/s | m/s |
| 0 - 10 | 10 | 1900 | 539 | 220 |
| 10 - 30 | 20 | 1900 | 980 | 400 |
| 30 - 50 | 20 | 1900 | 1347 | 550 |
| > 50 | | 1900 | 2450 | 1000 |

Table 4: Properties of a multilayered soil

## 5.4 Seismic wave polarization vs frame plan regularity

The case of a three-story one-span frame structure under seismic loading with different polarization is analyzed. Rectangular cross-section of beam elements is 30x90cm and the other properties are reported in Table 2 (frame T). All columns are in the same position. This frame is stiffer in $x$-direction, since columns have more inertia against horizontal actions in this direction. The same incident wave motion represented in Figure 4 is applied in $x$- and $y$-direction as 1-Component motion, and simultaneously as a 2-Component seismic loading. The soil reference shear strain is $\gamma_r = 0.35‰$.

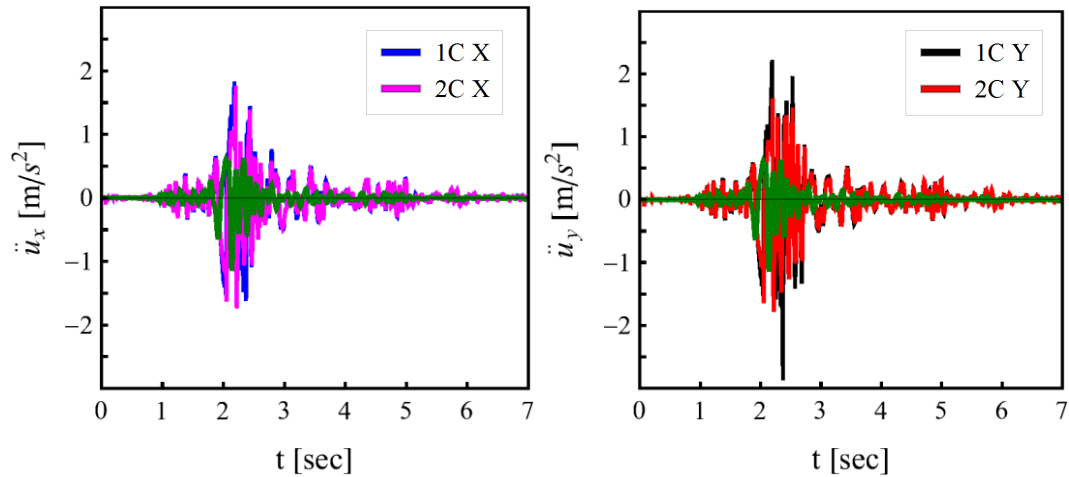

Figure 7: Soil-surface acceleration in the directions where the frame is stiffer (a) and softer (b), for 1- and 2-Component motion.

A reduction of PGA is observed from one to two components (from $\ddot{u}_{x\max} = 1.83\,\text{m}/\text{s}^2$ to $\ddot{u}_{x\max} = 1.76\,\text{m}/\text{s}^2$). The PGA is greater in $y$-direction (from $\ddot{u}_{y\max} = 2.86\,\text{m}/\text{s}^2$ to $\ddot{u}_{y\max} = 1.78\,\text{m}/\text{s}^2$), where the frame is softer, and lower in $x$- direction, where the frame is stiffer. Accelerations at the soil surface and at the top of the frame structure are lower in the direction where the structure is stiffer.

# 6 Conclusions

A model of one-directional three-component seismic wave propagation in a nonlinear multilayered soil profile is coupled with a multi-story multi-span frame model to consider the soil-structure interaction in a finite element scheme. Computation time is lower, compared with a 3D spatial discretization of soil and the boundary condition at the soil-bedrock interface is defined in only one node.
Modeling the simultaneous three-component wave propagation enables the analysis of the soil multiaxial stress state that reduces the soil strength and increases nonlinear effects. The variation of incident direction of seismic loading at the ground surface can be taken into account and the behavior of a frame structure shaken by a three-component earthquake can be observed.
A sensitivity analysis is carried out to define the appropriate soil column crosssectional area, allowing to appreciate the Soil-Structure-Interaction effects, without overestimate the influence of structure. Using a 1D soil model, the soil perceives the building as an elementary oscillator (one degree of freedom) characterized by its mass and stiffness. The number of storeys and spans modifies the total mass and stiffness and, consequently, the SSI effect, but the influence of an increasing floor area is not captured by a 1D soil model.
Local effects in the soil, dues to nonlinear behavior, impedance contrast between layers and wavefield polarization, are reproduced by the proposed model. The 3D model of the frame structure allows to estimate the response to a seismic acceleration at the base and to evaluate displacement components and internal forces. A linear behavior is assumed in this analysis for beams, but the proposed model is not dependent on the adopted constitutive relationship.
This 1D-3C-propagation-3D-frame model allows to confirm that a building is less stressed by a seismic wave having a frequency content close to the building principal frequency, if it is placed on a soil with a very different principal frequency, rather than in the case where soil and structure frequency content are close together. The acceleration at the soil surface and at the top of a building are reduced in the direction where the frame structure is stiffer against horizontal actions.
Further work would require a three-dimensional spatial discretization of soil, to take into account the influence of building floor area and effects of spatial variability in the seismic loading.

# Acknowledgements

This research has received funding from the European Union's Seventh Framework Program (FP7/2007-2013) under grant agreement n° 266638 (INTEGER project) as part of the PEPS Égalité Action (MISS project).
For the second author, a part of the research reported in this paper has been supported by the SEISM Paris Saclay Research Institute.